\documentclass[aps,twocolumn,pra,superscriptaddress,showpacs,tightenlines]{revtex4-1}
\usepackage{epsfig,graphicx,times}
\usepackage{amstext}
\usepackage{amsmath}
\usepackage{amssymb}
\usepackage{graphicx}
\usepackage{latexsym}
\usepackage{bm}
\usepackage[colorlinks,citecolor=blue,linkcolor=blue,hyperindex,CJKbookmarks,dvipdfm]{hyperref}

\begin{document}

\title{The energy level crossing behavior and quantum Fisher information in
a quantum well with spin-orbit coupling}
\author{Z. H. Wang}
\affiliation{Center for Quantum Sciences, Northeast Normal University,
Changchun 130117, China}
\affiliation{Beijing Computational Science Research Center, Beijing 100084,
China}
\author{Q. Zheng}
\affiliation{Beijing Computational Science Research Center, Beijing 100084,
China}
\affiliation{School of Mathematics and Computer Science, Guizhou Normal University, Guiyang 550001, China}
\author{Xiaoguang Wang}
\affiliation{Zhejiang Institute of Modern Physics, Department of Physics,
Zhejiang University, Hangzhou 310027, China}
\author{Yong Li}
\email{liyong@csrc.ac.cn}
\affiliation{Beijing Computational Science Research Center, Beijing 100084,
China}
\affiliation{Synergetic Innovation Center of Quantum Information and Quantum Physics, University of Science and Technology of China,
Hefei 230026, China}

\begin{abstract}
We study the energy level crossing behavior in two-dimensional quantum well
with the Rashba and Dresselhaus spin-orbit couplings (SOCs). By mapping the
SOC Hamiltonian onto an anisotropic Rabi model, we obtain the approximate
ground state and its quantum Fisher information (QFI) via performing a
unitary transformation. We find that the energy level crossing can occur in
the quantum well system within the available parameters rather than in
cavity and circuit quantum eletrodynamics systems. Futhermore, the influence
of two kinds of SOCs on the QFI is investigated and an intuitive explanation
from the viewpoint of the stationary perturbation theory is given.
\end{abstract}

\date{\today }
\maketitle

\section{Introduction}

In semiconductor physics, the spin-orbit coupling (SOC), which is available
to generate the so-called spin-orbit qubit~\cite{SN}, provides a useful
approach to manipulate the spin by electric field instead of magnetic field~%
\cite{RL}, and is widely studied in the field of both spintronics~\cite{IZ}
and quantum information processing~\cite{DD}. In the low dimensional
semiconductor, there exist two types of SOCs, that is the Rashba SOC which
comes from the structure inversion~\cite{Rash} and the Dresselhaus SOC which
comes from the bulk-inversion asymmetry~\cite{Dress}. In general cases, the
two types of SOCs coexist in a material~\cite{OV}.

{The spin properties of the electron(s) in semiconductor have been studied widely
and it shows that some novel features emerge when the SOC is present.
Among the various properties, the ones for the ground state play crucial
roles.}  In this paper, we study the ground state of the electron in a semiconductor
quantum well, which is subject to the Rashba and Dresselhaus SOCs as well as
a perpendicular magnetic field. The Hamiltonian in this two-dimensional
structure can be mapped onto a Hamiltonian describes a qubit interacting
with a single bosonic mode, where the spin degree of freedom of the electron
serves as the qubit and the orbit degree of freedom serves as the bosonic
mode~\cite{loss1,SD}. Furthermore, the Rashba SOC contributes to the
rotating wave interaction and the Dresselhaus SOC contributes to the
anti-rotating wave interaction. When the strengths and/or the phases of the
two types of SOCs are not equal to each other (this is the usual case in
realistic material), the mapped Hamiltonian is actually an anisotropic Rabi
model~\cite{qt} in quantum optics.

With the available parameters in quantum well systems, it will undergo the
energy level crossing between the ground and first excited states as the
increase of SOC strength. {This kind of energy level crossing will
induce a large entanglement for the ground state, and have some potential
applications in quantum information processing. Also, the steady state
of the system when the dissipation is present is also affected greatly by the
energy level crossing.} Although the same form of Hamiltonian (i.e., anisotropic Rabi
Hamiltonian) can also be achieved in cavity and circuit quantum electrodynamics (QED) systems, such a crossing would not occur since the related coupling between the bosonic mode and the qubit is too weak. In this paper, we analytically give the crossing
strength of Rashba SOC in which the energy level crossing occurs when the
Dresselhaus SOC is absent. Furthermore, we study the crossing phenomenon
numerically when both of the two kinds of SOCs are present.

The energy level crossing behavior in our system is similar to the
superradiant quantum phase transition in the Dicke model~\cite{Dicke}, where
the quantum properties (e.g., the expectation of photon number in ground
state) are subject to abrupt changes when the coupling strength between the
atoms and field reaches its critical value~\cite{nagy}. Recently, it is
found that the quantum Fisher information (QFI) is a sensitive probe to the
superradiant quantum phase transition in the Dicke model~\cite{jin1}. This
inspires us to investigate the relation between the QFI and energy level
crossing in our system.  {The QFI, as a key quantity in quantum estimation theory,
is introduced by extending the classical Fisher information to quantum regime, and can characterize the sensitivity of a state with respect to the change of a parameter. The
QFI is also related to quantum clone~\cite{yaoyao14}, and quantum Zeno dynamics~\cite{Smerzi12}.} In our system, we find that there exists an abrupt change in the
QFI at the crossing point, so that the QFI can be regarded as a signature of
the energy level crossing behavior in quantum well system. Furthermore, the
QFI increases with the increase of the Dresselhaus SOC.
As the increase of Rashba SOC strength, the QFI nearly remains before the
crossing but decreases monotonously after it. Actually, the QFI has a close
connection with the entanglement~\cite{LP}, and can be used to detect the
entanglement~\cite{LSL}, so our results can be explained from the viewpoint
of the entanglement and intuitively understood based on the stationary
perturbation theory.

The rest of the paper is organized as follows. In Sec.~\ref{model}, we
introduce the model under consideration and map the SOC Hamiltonian onto an
anisotropic Rabi model in quantum optics. Based on the mapped Hamiltonian,
we study in Sec.~\ref{crossing} the energy level crossing behavior when the
Dresselhaus SOC is on and off respectively. In Sec.~\ref{result}, we show
that the QFI of the ground state witnesses the energy level crossing, and
analyse its dependence on the Rashba and Dresselhaus SOCs before and after
the energy level crossing occurs. In Sec.~\ref{summary}, we give a brief
conclusion.

\section{System and Hamiltonian}

\label{model}

We consider an electron with mass $m_{0}$ and effective mass $m$ moving in a
two-dimensional $xy$ plane, which is provided by a semiconductor quantum
well. The electron is subject to the Rashba and Dresselhaus SOCs, and a
static magnetic field in positive $z$ direction $\vec{B}=B\vec{e_{z}}=\nabla
\times \vec{A}$. The Hamiltonian of the system is written as~\cite{loss1}
\begin{equation}
H=\frac{1}{2m}(\Pi _{x}^{2}+\Pi _{y}^{2})+\frac{1}{2}g\mu _{B}B\sigma
_{z}+H_{\mathrm{so}},  \label{oH}
\end{equation}%
where $\Pi _{x}$ ($\Pi _{y}$) is the x- (y-) direction component of the
canonical momentum $\vec{\Pi}\equiv \vec{p}+e\vec{A}$ with $\vec{p}$ the
mechanical momentum and $\vec{A}$ the vector potential. $g$ is the Lande
factor, and $\mu _{B}=e\hbar /2m_{0}$ is the Bohr magneton, $\sigma _{x,y,z}$
are the Pauli operators. Here, $\hbar $ is the Plank constant and $e=+|e|$
is the electronic charge.

The last term in Hamiltonian~(\ref{oH}), representing the SOCs, can be
divided into two terms $H_{\mathrm{so}}=H_{R}+H_{D}$, where
\begin{eqnarray}
H_{R} & = & \alpha(\Pi_{x}\sigma_{y}-\Pi_{y}\sigma_{x}), \\
H_{D} & = & \beta(\Pi_{x}\sigma_{x}-\Pi_{y}\sigma_{y}).
\end{eqnarray}
The Hamiltonian $H_R$ and $H_D$ represent the Rashba~\cite{Rash} and
Dresselhaus SOCs~\cite{Dress} term, respectively. $\alpha$ and $\beta$,
which are in units of velocity, describe the related strengths of the two
types of SOCs and are determined by the geometry of the heterostructure and
the external electric field across the field, respectively~\cite{Rash2}.

Since we consider that $\vec{B}$ is along the positive direction of $z$
axis, it is natural to choose the vector potential as $\vec{A}=(-y,x,0)/2$.
By defining the operator~\cite{loss1}
\begin{eqnarray}
b & = & \frac{1}{\sqrt{2\hbar eB}}(\Pi_{x}-i\Pi_{y}),
\end{eqnarray}
it is easy to verify that $[b,b^{\dagger}]=1$, so $b$ ($b^\dagger$) can be
regarded as a bosonic annihilation (creation) operator.

In terms of $b$ and $b^{\dagger }$, the Hamiltonian can be re-written as
\begin{equation}
H=E_{b}b^{\dagger }b+\frac{E_{a}}{2}\sigma _{z}+(\frac{\lambda _{1}}{2}%
b^{\dagger }\sigma _{-}+\frac{\lambda _{2}}{2}b\sigma _{-}+\mathrm{h.c.}),
\label{HO}
\end{equation}

where $\sigma _{\pm }=\sigma _{x}\pm i\sigma _{y}$, and the parameters are
calculated as

\begin{subequations}
\begin{eqnarray}
E_{b} &=&\frac{\hbar eB}{m},\,\ E_{a}=\frac{\hbar geB}{2m_{0}}, \\
\lambda _{1} &=&i\alpha \sqrt{2\hbar eB},\,\ \lambda _{2}=\beta \sqrt{2\hbar
eB}.
\end{eqnarray}
\label{para}
\end{subequations}

Thus, we have mapped the Hamiltonian in quantum well with SOCs onto a
standard anisotropic Rabi model~\cite{qt,LTS,zhang,ymin,AB} which describes
the interaction between a qubit and a single bosonic mode. Here the spin and
orbit degrees of freedom serve as the qubit and bosonic mode respectively.
In the language of quantum optics, the first two terms in Eq.~(\ref{HO}) are
the free terms of the boson mode with eig-energy $E_{b}$ and the qubit with
the transition energy $E_{a}$ respectively. The first term as well as its
hermitian conjugate in the braket in Eq.~(\ref{HO}) 
represents the rotating-wave coupling with strength $\lambda _{1}$ and the
second 
term as well as its hermitian conjugate represents the anti-rotating
coupling with strength $\lambda _{2}$, the relative phase between these two
coupling strengths is $\pi /2$ [see Eq.~\ref{para}(b)]. Actually, such a
kind of mapping from spintronics to quantum optics can also be performed
when an additional harmonic potential is added to confine the spatial
movement of the electron~\cite{SD,nzhao}.

In our system, both of the bare energies of the qubit and bosonic mode as
well as their coupling strength can be adjusted by changing the amplitude of
the external magnetic field, so that the coupling strength can be either
smaller or even (much) larger than the bare energies, this fact will lead to
some intrinsic phenomena, such as the energy level crossing, which will be
studied in detail in next section.

\section{Energy level crossing}

\label{crossing}

Based on the mapped anisotropic Rabi Hamiltonian in the above section, we
will discuss the energy level crossing~\cite{SA,DV} in this section.
Firstly, we give the crossing point without the Dresselhaus SOC
analytically, then the crossing behavior for the full Hamiltonian is
discussed numerically.

\subsection{Without Dresselhaus SOC}

When the Dresselhaus SOC is absent ($\beta =0$, then $\lambda _{2}=0$), the
mapped Hamiltonian reduces to the exact Jaynes-Cummings (JC) Hamiltonian
where the excitation number is conserved. The eigen-state without excitation
is $|E_{0}\rangle =|0;g\rangle :=|0\rangle _{o}\otimes |g\rangle _{s}$,
which represents that the orbit degree of freedom is in the bosonic vacuum
state and the spin degree of freedom is in its ground state (actually, is
the spin-down state because the magnetic field is in $+z$ direction in our
consideration). The corresponding eigen-energy is $E_{0}=-E_{a}/2$. In the
subspace with only one excitation, the pair of dressed states are

\begin{eqnarray}
|1+\rangle &=&-\cos \frac{\theta }{2}|1;g\rangle +i\sin \frac{\theta }{2}%
|0;e\rangle , \\
|1-\rangle &=&\cos \frac{\theta }{2}|0;e\rangle -i\sin \frac{\theta }{2}%
|1;g\rangle ,  \label{1n}
\end{eqnarray}%
and the corresponding eigen-energies are
\begin{equation}
E_{1\pm }=\frac{E_{b}}{2}\pm \frac{1}{2}\sqrt{\Delta ^{2}+|\lambda _{1}|^{2}}%
.
\end{equation}%
In the above equations, we have defined $\Delta :=E_{b}-E_{a}$, and
\begin{equation}
\tan \theta =|\lambda _{1}|/\Delta .  \label{eq:theta}
\end{equation}

Using the above results, it is shown that the ground state of the system is
either the separated state $|E_{0}\rangle $ when $E_{1-}>E_{0}$, or the
entangled state $|1-\rangle $ when $E_{1-}<E_{0}$. A simple calculation
gives the crossing Rashba SOC strength ($\lambda _{1}^{c}$) which separates
the entangled from unentangled (separated) ground state as
\begin{equation}
|\lambda _{1}^{c}|=2\sqrt{E_{a}E_{b}}.
\end{equation}

\subsection{With Dresselhaus SOC}

On the other hand, when the Dresselhaus SOC is present, the mapped
Hamiltonian yields an anisotropic Rabi Hamiltonian, in which the
rotating-wave term and the anti-rotating-wave term coexist. In this case,
the conservation of the excitation is broken, that is, $[\sigma
_{z}/2+b^{\dagger }b,H]\neq 0$. The analytical solution of quantum Rabi
model ($\lambda _{1}=\lambda _{2}$) was originally obtained by Braak~\cite%
{braak} and was developed to the case of anisotropic Rabi model ($\lambda
_{1}\neq \lambda _{2}$)~\cite{qt}. Their results however are based on a
composite transcendental function defined by power series, and are difficult
to extract the fundamental physics.

To deal with the anti-rotating-wave coupling term approximately, we now
resort to a unitary transformation to the Hamiltonian $H$~\cite%
{zhiguo,GJ,qingai,LTS},
\begin{equation}
H^{\prime }=e^{S}He^{-S}
\end{equation}%
with
\begin{equation}
S=(\xi b^{\dagger }-\xi ^{\ast }b)\sigma _{x},
\end{equation}%
where the parameter $\xi $ is to be determined.

Following the similar scheme in Ref.~\cite{LTS}, the transformed Hamiltonian
is obtained as $H^{\prime }=H_{a}+H_{b}+H_{c}$ where
\begin{eqnarray}
H_{a} &=&\frac{\tilde{E_{a}}}{2}\sigma _{z}+(E_{b}-\tilde{E_{b}}\sigma
_{z})b^{\dagger }b+E, \\
H_{b} &=&\frac{[(\lambda _{1}+\lambda _{2}-4\xi E_{b})b^{\dagger
}+h.c.]\sigma _{x}}{4}  \notag \\
&&-i\eta \sigma _{y}\{[E_{a}-\tilde{E_{c}}+\frac{\lambda
_{1}-\lambda _{2}}{4}]b^{\dagger }-h.c.\},  \label{eq:hb} \\
H_{c} &=&\frac{1}{2}(E_{a}-\tilde{E_{c}})(\hat{\kappa} _{1}\sigma
_{z}-i\hat{\kappa} _{2}\sigma _{y})+\hat{O}  \notag \\
&&+\frac{1}{4}(\hat{\kappa} _{2}\sigma _{z}-i\hat{\kappa}_1\sigma _{y}
)[(\lambda _{1}-\lambda _{2})b^{\dagger }-h.c.]  \label{ncc}
\end{eqnarray}%
with $\eta =e^{-2|\xi |^{2}}$,  $\tilde{E}_{a}:=E_{a}\eta -\tilde{E}_{b}$, $%
\tilde{E}_{b}:=\mathrm{Re}[\xi ^{\ast }(\lambda _{1}-\lambda _{2})]$, and $%
\tilde{E_{c}}:=i\mathrm{Im}[\xi ^{\ast }(\lambda _{1}-\lambda _{2})]$. Here
$\hat{O}:=(\lambda _{1}-\lambda _{2})\xi \eta b^{\dagger 2}\sigma _{z}/2+h.c.
$ {corresponds to the ``two-photon'' processing.} And we have also defined $\hat{\kappa}_{1}:=\cosh [2(\xi b^{\dagger }-\xi
^{\ast }b)]-\eta $ and $\hat{\kappa}_{2}:=\sinh [2(\xi b^{\dagger }-\xi
^{\ast }b)]-2(\xi b^{\dagger }-\xi ^{\ast }b)$, which are in the order of $%
|\xi |^{2}$ and $|\xi |^{3}$, respectively.

When $|\xi |\ll 1$ (which is valid as shown in what follows), we will
neglect $H_{c}$ and then $H^{\prime }=H_{a}+H_{b}$. For further simplicity,
we can properly choose $\xi $ to eliminate the anti-rotating-wave terms in $%
H_{b}$, for which $\xi $ satisfies

\begin{equation}
\eta^{-1}(\frac{\lambda _{1}+\lambda _{2}}{4}-E_{b}\xi)=E_{a}\xi +\frac{1}{4}%
(\lambda _{1}-\lambda _{2}) - \xi \tilde{E_c} ,
\label{eq:condition}
\end{equation}%
and then
\begin{equation}
H_{b}=(\frac{\lambda _{1}+\lambda _{2}}{2}-2E_{b}\xi )b^{\dagger }\sigma
_{-}+h.c..
\end{equation}%
Thus, the approximate Hamiltonian $H^{\prime }=H_{a}+H_{b}$ can be solved
exactly.

\begin{figure}[tbp]
\begin{centering}
\includegraphics[width=8 cm]{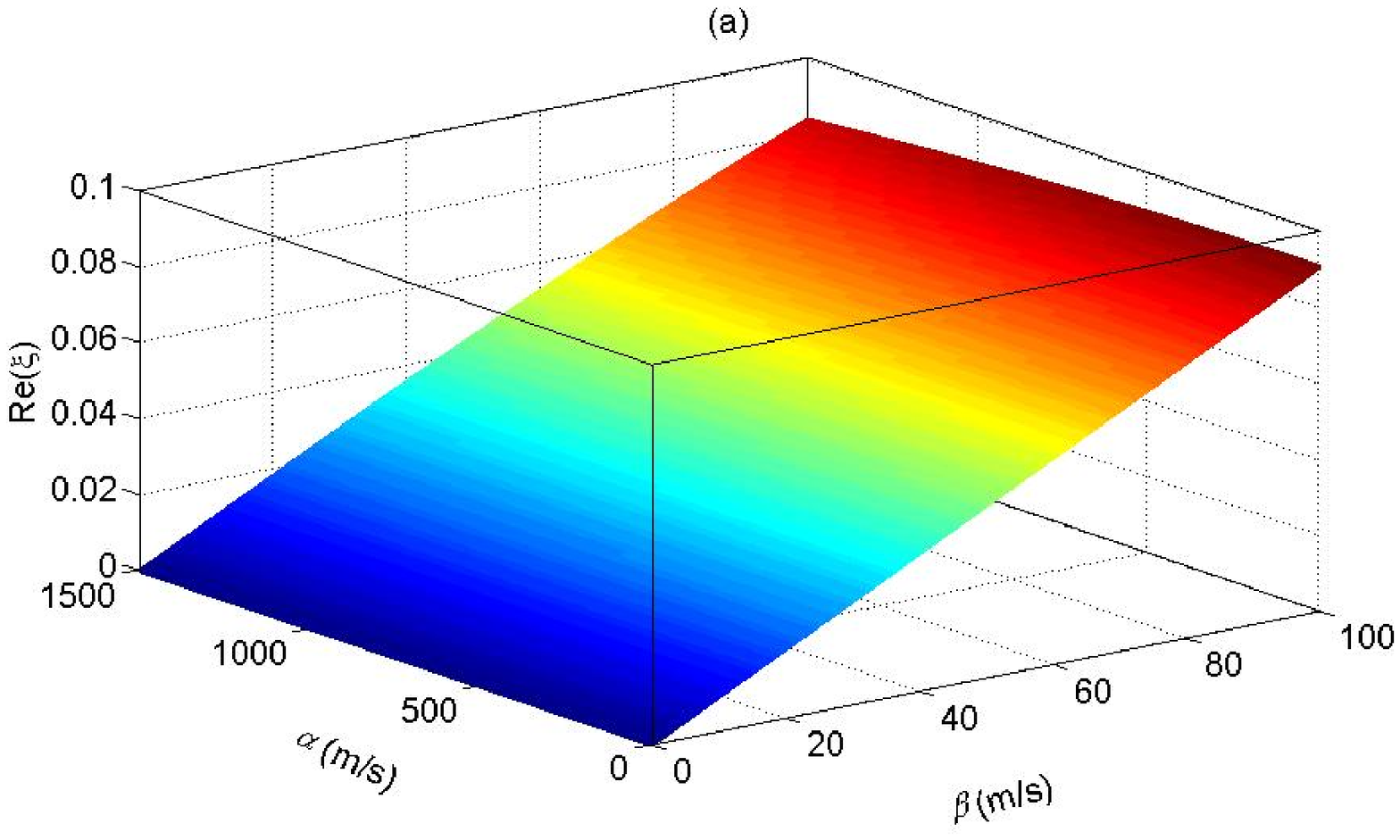}
\par \end{centering}
\begin{centering}
\includegraphics[width=8 cm]{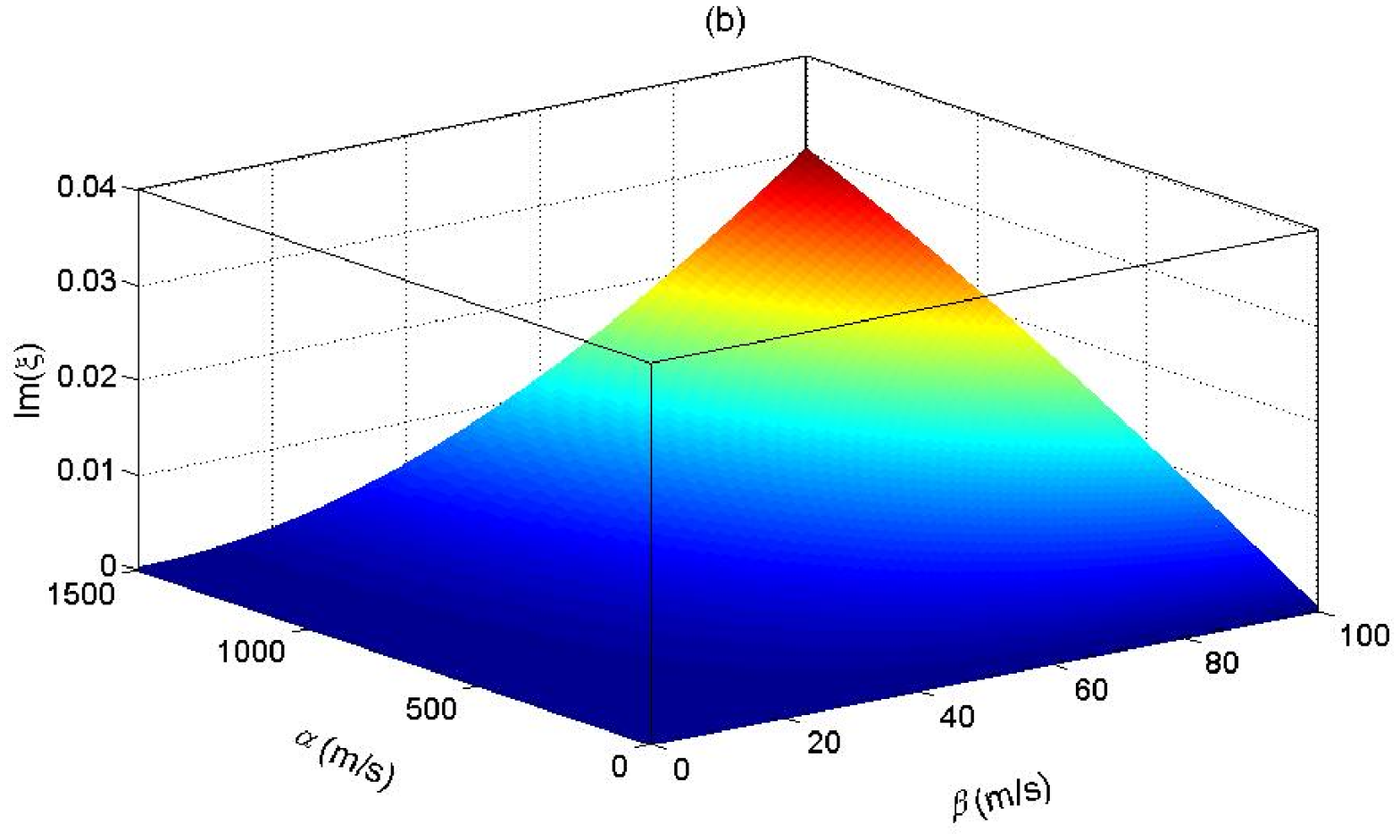}
\par \end{centering}
\caption{(Color online) The real and imaginary parts of $\protect\xi$ as
functions of $\protect\alpha$ and $\protect\beta$. The parameters are chosen
as $g=1.52$, $m_0=9\times10^{-31}$ kg, $m=0.15m_0$, $B=0.01$T. Under these
parameters, we will have $E_a/\hbar\approx1.35$ GHz, $E _b/\hbar\approx1.70$
GHz, and $(|\protect\lambda_1/\hbar|,|\protect\lambda _2/\hbar|)\approx5.52%
\times(\protect\alpha,\protect\beta)$ MHz. }
\label{xi}
\end{figure}

We note that, when both $\lambda_1$ and $\lambda_2$ are real numbers, $\xi$
is also real, then our transformed Hamiltonian and the equation $\xi$
satisfies coincide exactly with those in Ref.~\cite{LTS}. However, as shown
in Sec.~\ref{model} [see Eq.~(\ref{para})], here $\lambda_1$ is a pure
imaginary number and $\lambda_2$ is a real number, so that $\xi$ is a
complex number. We numerically solve Eq.~(\ref{eq:condition}), and plot the
real and imaginary parts of $\xi$ in Fig.~\ref{xi} as functions of $\alpha$
and $\beta$ for $B=0.01$\,T. It is obvious that $|\xi|$ is indeed much
smaller than $1$, so that we can safely neglect the effect of $H_c$ which is
at least in the order of $|\xi|^2$.

It is obvious that the approximate Hamiltonian $H\mathcal{^{\prime }}$ has a
similar form as the JC Hamiltonian, and the energy spectrum in the subspace
of zero- and one-excitation are obtained as
\begin{eqnarray}
E_{0}^{d} &=&-\frac{\tilde{E}_{a}}{2}+E, \\
E_{1\pm }^{d} &=&E+\frac{E_{b}+\tilde{E}_{b}}{2}\pm \sqrt{|\tilde{g}|^{2}+%
\tilde{\Delta}^{2}},
\end{eqnarray}%
where
\begin{subequations}
\begin{eqnarray}
\tilde{g} &=&\frac{\lambda _{1}+\lambda _{2}}{2}-2E_{b}\xi , \\
E &=& E_b|\xi|^{2}-\frac{\mathrm{Re}[(\lambda _{1}+\lambda _{2})\xi]} {2}, \\
\tilde{\Delta} &=&\frac{E_{b}+\tilde{E}_{b}-\tilde{E}_{a}}{2}.
\end{eqnarray}

The energy level crossing then occurs when $E_{1-}^{d}=E_{0}^{d}$. In Fig.~\ref{gap},
we plot the energy gap $\Delta E^{d}:=E_{1-}^{d}-E_{0}^{d}$ as a
function of $\alpha$ and $\beta$.

\begin{figure}[tbp]
\begin{centering}
\includegraphics[width=8 cm]{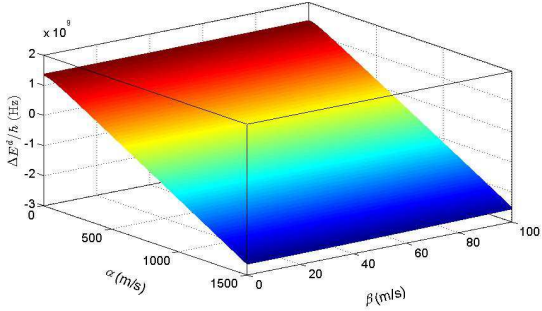}
\par \end{centering}
\caption{(Color online) The energy gap as a function of $\protect\alpha$ and
$\protect\beta$. The parameters are same as those in Fig.~\protect\ref{xi}.}
\label{gap}
\end{figure}

As shown in Fig.~{\ref{gap}}, for small $\alpha$, $\Delta E^{d}>0$, and the
ground state is
\end{subequations}
\begin{eqnarray}
|G_1\rangle & = & e^{-S}|0;g\rangle  \notag \\
& = & \frac{1}{\sqrt{2}}(|-\xi;+\rangle-|\xi;-\rangle),  \label{ground}
\end{eqnarray}
where $|\pm\rangle$ are the eigen states of Pauli operator $\sigma_x$
satisfying $\sigma_x|\pm\rangle=\pm|\pm\rangle$, and $|\pm\xi\rangle$ are
the bosonic coherent states with amplitudes $\pm\xi$.

As the increase of $\alpha$, the energy level crossing occurs, that is, $%
\Delta E^{d}$ becomes negative, and the ground state becomes
\begin{eqnarray}  \label{ground2}
|G_{2}\rangle & = & e^{-S}[\cos\frac{\theta_{d}}{2}|0;e\rangle-\sin\frac{%
\theta_{d}}{2}e^{i\phi}|1;g\rangle]  \notag \\
& = & e^{-(\xi b^{\dagger}-\xi^{*}b)\sigma_{x}}[\cos\frac{\theta_{d}}{2}%
|0;e\rangle -\sin\frac{\theta_{d}}{2}e^{i\phi}|1;g\rangle],  \notag \\
\end{eqnarray}
where
\begin{equation}
\tan\theta_{d}=\frac{|\tilde{g}|}{\tilde{\Delta}},\,\tan\phi=\frac{\mathrm{Im%
}[\tilde{g}]}{\mathrm{Re}[\tilde{g}]}.
\end{equation}

In the end of this section, we would like to emphasize the following point.
Besides the quantum well system as discussed in this paper, the anisotropic
Rabi model can be realized in various systems, e.g., in the cavity or
circuit QED systems. In a typical cavity QED system, in which the atom
interacts with the optical cavity mode, the frequencies of the atomic
transition and the cavity mode are of the order of $10^{4}-10^{5}$ GHz, and
the coupling strength reaches hundreds of MHz~\cite{AL}. In a typical
circuit QED system, where the artificial atom (superconducting qubit)
couples to the transmission line resonator, the frequencies of the qubit and
the resonator are about several GHz, and the coupling strength can be
realized by hundreds of MHz~\cite{IC,KV}. In these two kinds of systems,
which motivate many research interests during the past decades, the energy
level crossing hardly occurs since the coupling strength is not strong
enough. However, the energy level crossing can be available in the realistic
quantum well material. Taking the AlAs material as an example, the Lande
factor is $g=1.52$, the mass of electron is $m_{0}=9\times 10^{-31}$%
\thinspace kg, and the effective mass is $m=0.15m_{0}$~\cite{MV}. When the
quantum well is subject to a magnetic field $B=0.01$\thinspace T in $+z$
direction, we will have $E_{a}/\hbar \approx 1.35$\thinspace GHz, $%
E_{b}/\hbar \approx 1.70$\thinspace GHz, and $(|\lambda _{1}|/\hbar
,|\lambda _{2}|/\hbar )\approx 5.52\times (\alpha ,\beta )$\thinspace MHz.
When choosing the parameters $\alpha $ in the order of hundreds of m/s and $%
\beta $ in the order of tens of m/s, which can be achieved easily with
recent available experimental techniques~\cite{loss1}, the coupling strength
could be in the same order or even larger than the energies of $E_{a}$ and $%
E_{b}$. Therefore, the two-dimensional quantum well system provides a
promising platform to simulate the energy level crossing behavior and
related phenomenon.

\section{Quantum Fisher information of the ground state}

\label{result}

In the above section, we have depicted the energy level crossing behavior in
our system. {In this section, the QFI of the ground state is adopted to further characterize the level crossing behavior, and its dependence on $\alpha$ and $\beta$ is detailed studied.} We will also give an intuitive
explanation about the obtained results based on the stationary perturbation
theory.

{The so-called quantum Cram\'{e}r-Rao (CR) inequality, obtained by extending the classical CR inequality to
quantum probability and choosing the quantum
measurement procedure for any given quantum state to maximize the classical CR inequality,
gives a bound to the variance Var$(\hat{\varphi})$ of any
unbiased estimator $\hat{\varphi} $~\cite{SL},
\begin{equation}
\text{Var}(\hat{\varphi} )\geq \frac{1}{NF_{\varphi}},
\end{equation}%
where $F_{\varphi}$ is the QFI and $N$ is the number of independent measurements. A larger QFI
corresponds to a more accurate estimation to the parameter $\varphi$.
Moreover, the QFI is connected to the Bures distance~\cite{SL} through
\begin{equation}
D_{B}^2[\rho_{\varphi}, \rho_{\varphi+d\varphi}]= \frac{1}{4}F_{\varphi}d \varphi^2,
\end{equation}
where the Bures distance is defined as $ D_{B}[\rho, \sigma]:=[ 2(1-\mathrm{Tr}\sqrt{\rho^{1/2} \sigma \rho^{1/2}})]^{1/2}$.}

{For an arbitrary given quantum state, its QFI can be determined by the spectrum decomposition of the state.
Fortunately, for a pure quantum state given by
the wave function $|\psi \rangle $, its QFI with respect to the parameter $%
\varphi$ has a relatively simple form, given as~\cite{SL2,weizhong,YMZhang,Qzheng}}
\begin{equation}
F_{\varphi }=4(\langle \partial _{\varphi }\psi |\partial _{\varphi }\psi
\rangle -|\langle \psi |\partial _{\varphi }\psi \rangle |^{2}).
\label{fisher}
\end{equation}

\begin{figure}[tbp]
\begin{centering}
\includegraphics[width=8 cm]{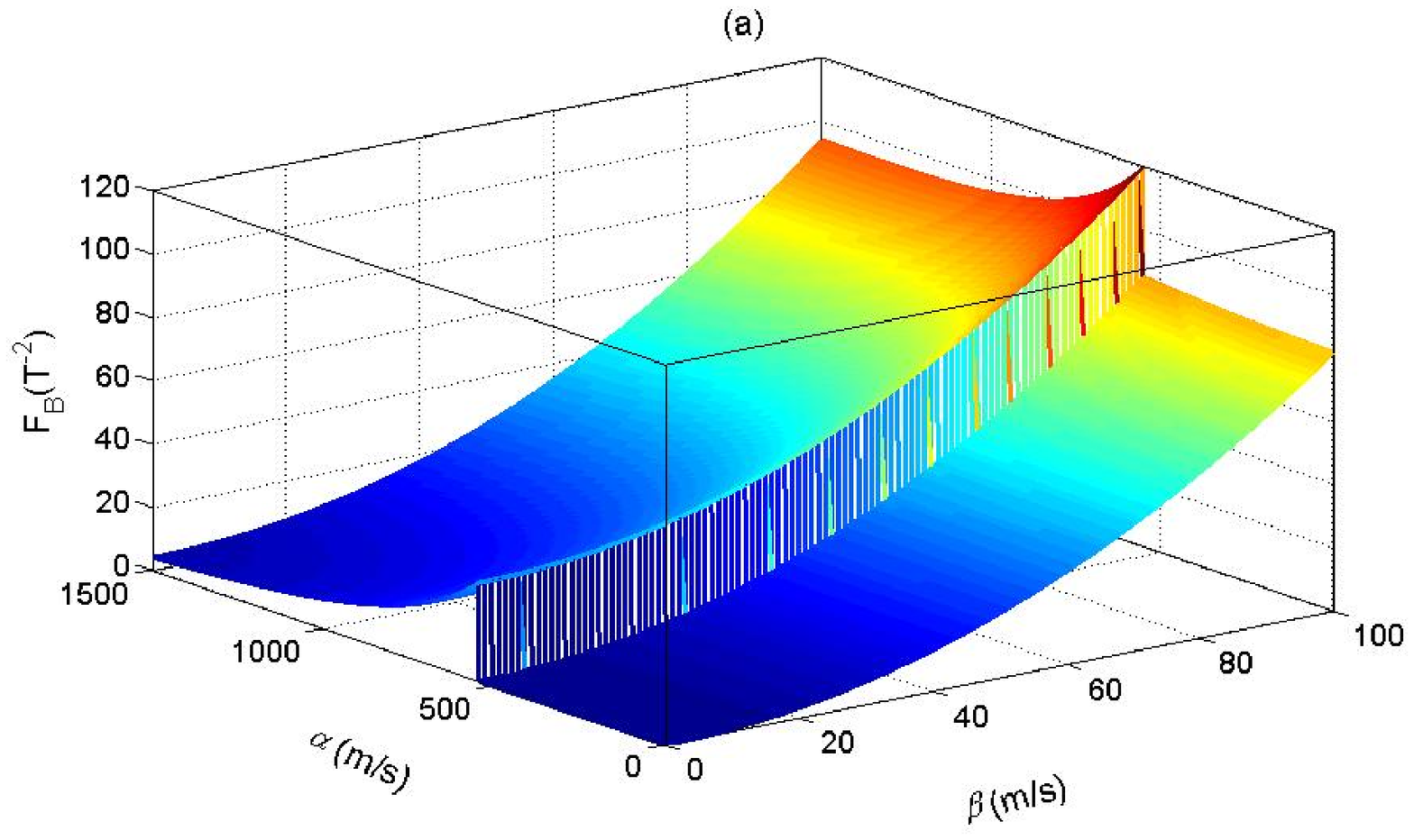}
\par \end{centering}
\begin{centering}
\includegraphics[width=8 cm]{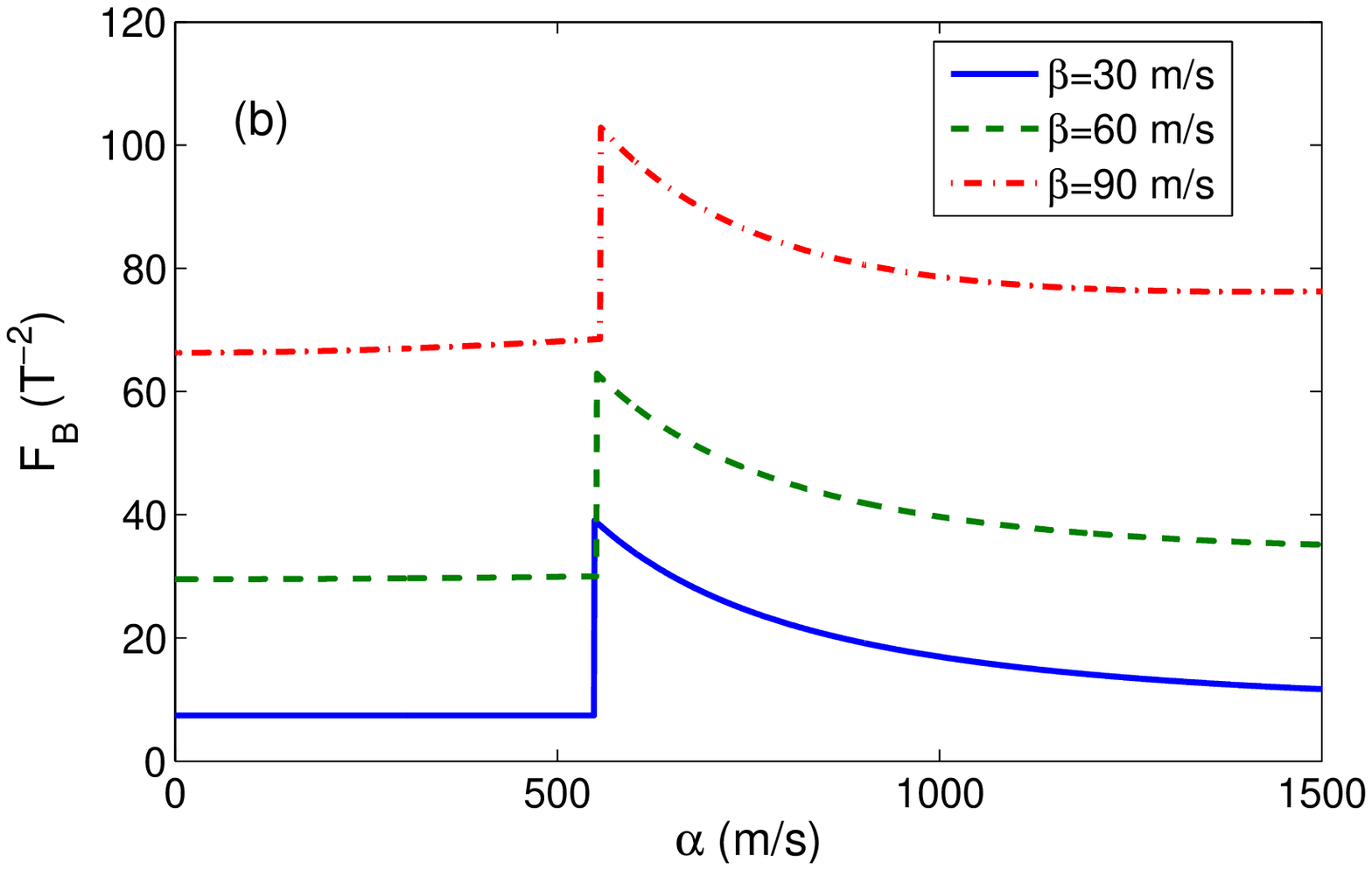}
\par \end{centering}
\begin{centering}
\includegraphics[width=8 cm]{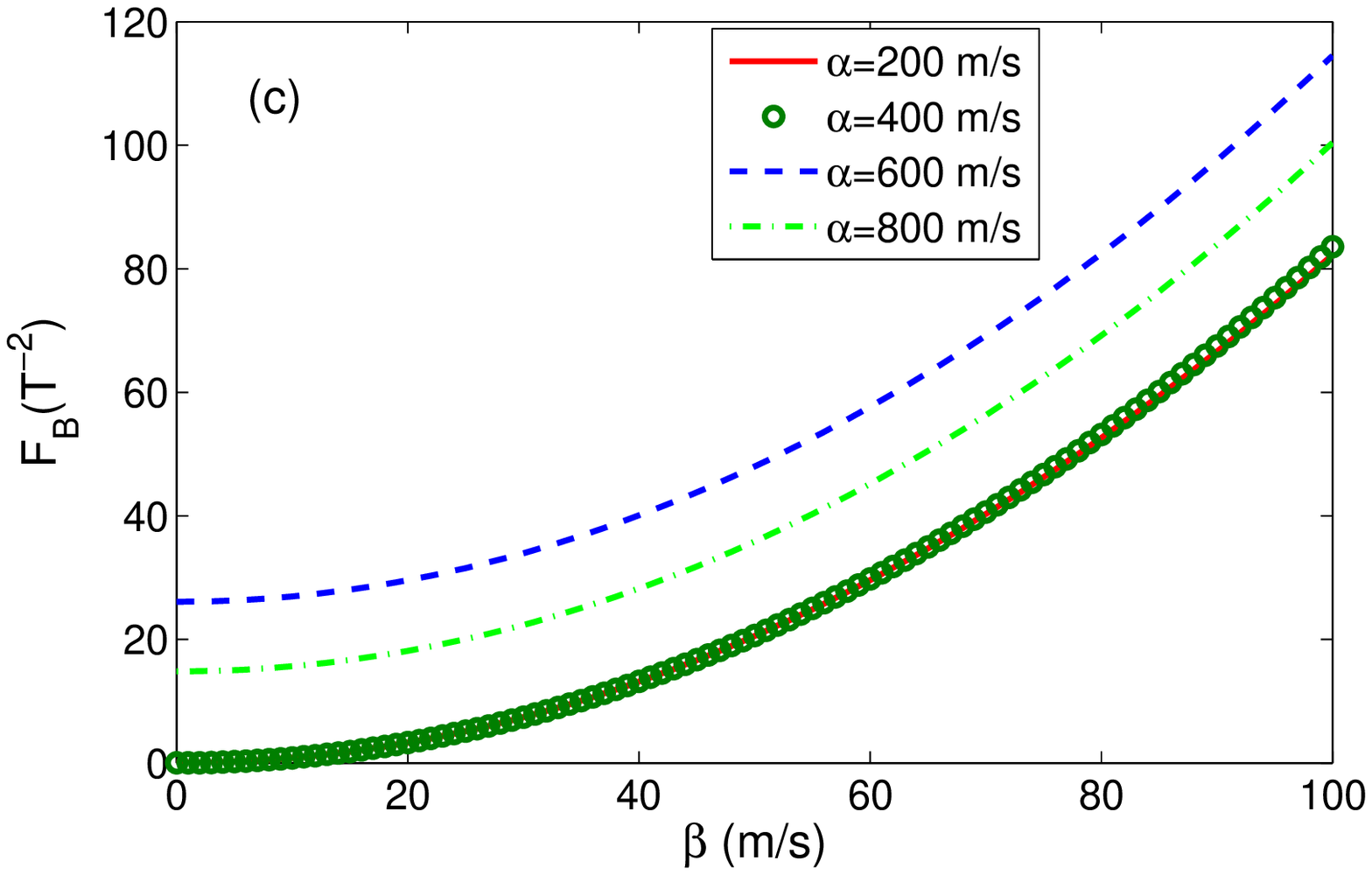}
\par \end{centering}
\caption{(Color online) (a) The QFI with respect to the external magnetic
field $B$, $F_B$, as a function of $\protect\alpha$ and $\protect\beta$. (b)
$F_B$ as a function of $\protect\alpha$ for different $\protect\beta$. (c) $%
F_B$ as a function of $\protect\beta$ for different $\protect\alpha$. The
other parameters are the same as those in Fig.~\protect\ref{xi}. }
\label{QFI}
\end{figure}

Before the energy level crossing occurs, that is $\Delta E^{d}>0$, the
ground state of the system with the presence of both the Rashba and
Dresselhaus SOCs is shown in Eq.~(\ref{ground}). After some straightforward
calculations, its QFI with respect to the external magnetic field $B$ is
given as
\begin{equation}
F_{B}=4[(\frac{\partial \mathrm{Re}{\xi }}{\partial B})^{2}+(\frac{\partial
\mathrm{Im}{\xi }}{\partial B})^{2}].
\end{equation}%

After the energy level occurs, that is $\Delta E^{d}<0$, the ground state is
given in Eq.~(\ref{ground2}), the expression of the QFI with respective to $%
B $ is tedious and we will only give the numerical results here.

In Fig.~\ref{QFI}(a), we plot the QFI as a function of $\alpha$ and $\beta$.
It obviously shows that the QFI undergoes a sudden change when the energy
level crossing occurs. Therefore, the QFI of the ground state can be
regarded as a witness to the energy level crossing behavior.

Furthermore, in Fig.~\ref{QFI}(b), we plot the QFI as a function of $\alpha $
for different values of $\beta $. On one hand, the QFI nearly keeps constant
when $\alpha $ approaches the crossing point from small values, and
decreases monotonously when $\alpha $ surpasses the crossing value $\alpha^{c}$ ($\alpha
^{c}\approx 550$ m/s within our chosen parameters and it corresponds to $|\lambda _{1}^{c}|/\hbar \approx
3$ GHz). On the other hand, a larger $\beta $ will lead to a larger QFI,
which implies a more precise measurement about the magnetic field. This
result is also demonstrated in Fig.~\ref{QFI}(c), where the QFI is plotted
as a function of $\beta $ for different values of $\alpha $. It shows that
the curves for $\alpha =200$\thinspace m/s and $\alpha =400$\thinspace m/s,
which are both below the crossing values, coincide with each other. As for
the values above the crossing point, we observe a decreasing behavior of QFI
as the increase of $\alpha $, for example, the QFI for $\alpha =600$ m/s is
larger than that for $\alpha =800$ m/s as shown in Fig.~\ref{QFI}(c).

The dependence of QFI on the strengths of the Rashba and Dresselhaus SOCs
can be explained from the viewpoint of the stationary perturbation theory
qualitatively as what follows. In our consideration, the strength of
Dresselhaus SOC is much weaker than that of the Rashba SOC and the bare
energy of spin/orbit degree of the freedom, so it can be regarded as a
perturbation. In this sense, the mapped Hamiltonian [Eq.~(\ref{HO})] can be
divided into $H=H_{0}+H_{I}$, where the un-perturbation part is
\begin{equation}
H_{0}=E_{b}b^{\dagger }b+\frac{E_{a}}{2}\sigma _{z}+(\frac{\lambda _{1}}{2}%
b^{\dagger }\sigma _{-}+\frac{\lambda _{1}^{\ast }}{2}\sigma _{+}b),
\end{equation}%
and the perturbation part is
\begin{equation}
H_{I}=(\frac{\lambda _{2}}{2}b\sigma _{-}+\frac{\lambda _{2}^{\ast }}{2}%
\sigma _{+}b^{\dagger }).
\end{equation}

For small $\alpha $ or $|\lambda _{1}|$, the ground state of $H_{0}$ is $%
|0;g\rangle $ which is independent of the field $B$ and yields a zero QFI.
The perturbation part, which is contributed from the Dresselhaus SOC, mixes
the state $|0;g\rangle $ with $|1;e\rangle $, yields an entangled ground
state and gives a non-zero $\beta $ ($|\lambda _{2}|$) dependent QFI. It is
obvious that the Dresselhaus SOC will enhance the entanglement, so that the
QFI also increases as $\beta $ becomes larger.

For large $\alpha $ or $|\lambda _{1}|$, the energy level crossing occurs,
and the ground state of $H_{0}$ becomes the wave function given in Eq.~(\ref%
{1n}), which is an entangled state, yields a non-zero QFI. Furthermore, the
entanglement decreases (increases) with the increase of $\alpha $ ($\beta $%
), and so the QFI behaves in a similar way.

\section{Conclusion}

\label{summary}

In this paper, we investigate the energy level crossing behavior and the QFI
of the ground state in the AlAs semiconductor quantum well. The Hamiltonian
of the system with the Rashba and Dresselhaus SOCs simultaneously is mapped
onto an anisotropic Rabi model in quantum optics. We find that although the
mapped Hamiltonian is similar to that in cavity and circuit QED systems, the
energy level crossing behavior only occurs in our current system with the
available parameters. As a probe of the energy level crossing in our system,
we discuss the QFI of the ground state and find that the QFI exhibits
different dependences on the strengths of the Rashba and Dresselhaus SOCs
and has a sudden jump when the crossing happens. Based on the stationary
perturbation theory, we give an intuitive explanation to the results.

\begin{acknowledgments}
We thank P. Zhang for his fruitful discussions. This work is supported the National Basic Research Program of China (under Grant No. 2014CB921403 and No. 2012CB921602) and by NSFC (under Grants No. 11404021, No. 11475146 and No. 11422437). Q. Zheng is supported by
NSFC (under Grant) No. 11365006 and Guizhou province science and technology innovation talent team (Grant No. (2015)4015).

\end{acknowledgments}

\end{document}